\documentclass[amsmath,amssymb,prb,preprintnumbers,superscriptaddress, twocolumn,showpacs]{revtex4}

\usepackage{graphicx}

\usepackage{dcolumn}
\usepackage{bm}         
\usepackage{amssymb}
\usepackage{amsmath}
\usepackage{ulem}     

\newcommand{\gapprox}{{\scriptscriptstyle\stackrel{>}{\sim}}}
\newcommand{\lapprox}{{\scriptscriptstyle\stackrel{<}{\sim}}}
\newcommand{\rb}[1]{\raisebox{1.5ex}[-1.5ex]{#1}}
\newcommand{\lb}[1]{\raisebox{-1.5ex}[-1.5ex]{#1}}

\begin{document}

\date{\today}


\title
{Optimizing the spin sensitivity of grain boundary junction nanoSQUIDs -- towards detection of small spin systems with single-spin resolution}

\author{R.~W\"{o}lbing}
\author{T.~Schwarz}
\author{B.~M\"{u}ller}
\author{J.~Nagel}
\author{M.~Kemmler}
\author{R.~Kleiner}
\author{D.~Koelle}
\affiliation{
Physikalisches Institut -- Experimentalphysik II and Center for Collective Quantum Phenomena in LISA$^+$,
Universit\"at T\"ubingen,
Auf der Morgenstelle 14,
D-72076 T\"ubingen, Germany}

\begin{abstract}

We present an optimization study of the spin sensitivity of nanoSQUIDs based on resistively shunted grain boundary Josephson junctions.
In addition the dc SQUIDs contain a narrow constriction onto which a small magnetic particle can be placed (with its magnetic moment in the plane of the SQUID loop and perpendicular to the grain boundary) for efficient coupling of its stray magnetic field to the SQUID loop.
The separation of the location of optimum coupling from the junctions allows for an independent optimization of the coupling factor $\phi_\mu$ and junction properties.
We present different methods for calculating $\phi_\mu$ (for a magnetic nanoparticle placed 10\,nm above the constriction) as a function of device geometry and show that those yield consistent results.
Furthermore, by numerical simulations we obtain a general expression for the dependence of the SQUID inductance on geometrical parameters of our devices, which allows to estimate their impact on the spectral density of flux noise $S_\Phi$ of the SQUIDs in the thermal white noise regime.
Our analysis of the dependence of $S_\Phi$ and $\phi_\mu$ on the geometric parameters of the SQUID layout yields a spin sensitivity $S_\mu^{1/2}=S_\Phi^{1/2}/\phi_\mu$ of a few $\mu_{\rm{B}}/\rm{Hz^{1/2}}$ ($\mu_B$ is the Bohr magneton) for optimized parameters, respecting technological constraints.
However, by comparison with experimentally realized devices we find significantly larger values for the measured white flux noise, as compared to our theoretical predictions.
Still, a spin sensitivity on the order of $10\,\mu_{\rm B}/\rm{Hz^{1/2}}$ for optimized devices seems to be realistic.
\end{abstract}



\maketitle

\section{Introduction}
\label{sec:Introduction}

Miniaturized direct current (dc) superconducting quantum interference devices (SQUIDs) with dimensions in the sub-micrometer range (nanoSQUIDs) are promising devices for the sensitive detection and investigation of small spin systems \cite{Wernsdorfer01}.
The basic idea behind this is to attach a small (nanometer-sized) magnetic particle directly to the SQUID and trace out magnetic hysteresis loops of the particle.
This shall be done by detecting the change of the stray magnetic field of the particle with magnetic moment $\bm\mu$ via the change of the magnetic flux $\Phi$ coupled to the SQUID loop \cite{Ketchen89,Wernsdorfer00,Bouchiat09}.
To meet the ultimate goal of detecting the flipping of only a few electron spins \cite{Gallop03}, the spin sensitivity $S_\mu^{1/2}=S_\Phi^{1/2}/\phi_\mu$ has to be optimized carefully via reducing the spectral density of flux noise $S_\Phi$ of the SQUID and increasing the coupling factor $\phi_\mu\equiv\Phi/\mu$ (with $\mu\equiv|\bm \mu|$).
$S_\Phi$ can be reduced by shrinking the size of the SQUID loop, and hence its inductance $L$, and $\phi_\mu$ can be increased by placing the particle on a narrow constriction inserted in the SQUID loop, which motivates the need to implement sub-micron SQUID structures.

Until now, the most common approach for the realization of nanoSQUIDs is to use constriction type Josephson junctions (cJJs) intersecting small SQUID loops (see e.g.~Ref.~[\onlinecite{Foley09}] published in a special issue on nanoSQUIDs and related articles therein).
Although impressive results have been achieved very recently for ultra-small SQUIDs based on Pb constrictions \cite{Vasyukov13}, the cJJ approach comes with several drawbacks:
Constriction type Josephson junctions often show hysteretic current-voltage characteristics (IVCs).
This hampers continuous operation of cJJ-based nanoSQUIDs, which however is required for the investigation of the magnetization dynamics of the sample under investigation.
Hence, more advanced readout-schemes are required for operating such devices.
We should note here, that very sensitive Nb thin film nanoSQUIDs based on cJJs, resistively shunted with a thin W layer, have been realized \cite{Hao08}.
However, in this case, the devices show optimum performance only in a narrow range of temperature $T$ not too far below the transition temperature $T_{\rm c}$ of Nb, which makes them less interesting for applications.
Also, the noise properties of cJJs are not well understood and hence hard to optimize.
And, finally, the magnetic particles have to be placed close to the cJJs to achieve optimum coupling.
However, this means that the junction properties and the coupling factor $\phi_\mu$ cannot be optimized independently, which hampers a careful optimization of the spin sensitivity.

With respect to the application of nanoSQUIDs for the detection of the magnetization reversal of nanomagnets, the most interesting regime of operation is at $T\approx 1\,$K and below and at very high magnetic fields in the tesla range \cite{Wernsdorfer01}.
It has been demonstrated that Nb thin film nanoSQUIDs based on constriction type junctions can be operated in impressive background fields up to 7\,T \cite{Chen10}.
However, the upper critical field $B_{\rm c2}$ of typical Nb thin films ($\sim 1\,$T) requires to use very thin Nb films with thicknesses of only a few nm, i.e.~well below the London penetration depth $\lambda_{\rm L}$ of the Nb films, if such SQUIDs shall be operated in tesla fields.
This leads to a large kinetic inductance contribution to the SQUID inductance, and hence a large flux noise of such SQUIDs, which does not allow to use the huge potential for the realization of ultralow-noise nanoSQUIDs.
We note that ultralow noise values have been achieved for ultra-small SQUIDs based on Pb cJJs up to $\sim 1$\,T, where the high-field operation was presumably also limited by $B_{\rm c2}$ \cite{Vasyukov13}.

To circumvent the above mentioned drawbacks, we recently started to develop dc nanoSQUIDs based on $c$-axis oriented YBa$_2$Cu$_3$O$_7$ (YBCO) thin films with submicron wide bicrystal grain boundary Josephson junctions (GBJs) \cite{Nagel11}.
Due to the huge upper critical field of YBCO, such SQUIDs can be realized with film thicknesses on the order of $\lambda_{\rm L}$ and above and operated in tesla fields.
Furthermore, due to the large critical current densities of the YBCO GBJs (several mA/$\mu$m$^2$ at $T=4.2\,$K and below for a grain boundary misorientation angle of 24$^\circ$) submicron junctions still yield reasonably large values of the critical current $I_0$.
To achieve non-hysteretic IVCs, the GBJs are shunted by a thin Au film.
Due to the fact that the barrier of the GBJs is oriented perpendicular to the YBCO thin film plane, it is possible to apply tesla magnetic fields in the plane of the film, without a significant reduction of $I_0$ \cite{Schwarz13}.
And finally, by implementing an additional narrow constriction (which can be much narrower than the GBJs) in the SQUID loop, the optimization of the coupling factor for a nanoparticle placed on top of the constriction is possible without affecting the junction properties.

Here, we present a detailed optimization study of the spin sensitivity of such grain boundary junction nanoSQUIDs by analyzing the dependence of the flux noise $S_\Phi$ and the coupling factor $\phi_\mu$ on the geometry of our devices.
We find that for an optimized SQUID geometry a continuous detection of magnetic moments down to a spin sensitivity $S_\mu^{1/2}$ of a few $\mu_{\rm B}/\rm{Hz^{1/2}}$ ($\mu_{\rm B}$ is the Bohr magneton) is feasible if a magnetic particle is placed 10\,nm above the center of the constriction, with its magnetic moment oriented in the plane of the SQUID loop and perpendicular to the grain boundary.

\section{nanoSQUID design}
\label{sec:nanoSQUIDdesign}

\begin{figure}[t]
\includegraphics[width=8cm]{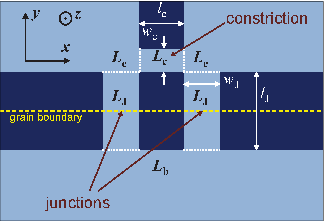}
\caption{Schematic view of the nanoSQUID layout, divided (by white dotted lines) into the constriction (inductance $L_{\rm c}$, length $l_{\rm c}$, width $w_{\rm c}$), two corners (each with inductance $L_{\rm e}$), the two junctions (each with inductance $L_{\rm J}$, length $l_{\rm J}$, width $w_{\rm J}$) and the bottom part (inductance $L_{\rm b}$).}
\label{Fig:layout}
\end{figure}

The layout of the nanoSQUID (top view) is shown in Fig.~\ref{Fig:layout}.
The SQUID structure is patterned in a YBCO thin film of thickness $d$, covered by a thin Au film with thickness $d_{\rm Au}$.
The two bridges straddling the grain boundary have a width $w_{\rm J}$ and length $l_{\rm J}$.
The upper part of the SQUID loop contains a constriction of width $w_{\rm c}$ and length $l_{\rm c}$.
An applied bias current $I_{\rm b}$ is flowing from top to bottom across the two GBJs.
A small magnetic particle can be placed on top of the constriction, and an in-plane magnetic field (perpendicular to the grain boundary, i.e.~along the $y$-direction) can be applied without significant suppression of the critical current $I_0$ of the two GBJs.

Optimizing the SQUID for spin sensitivity means to minimize the ratio $S_\Phi/\phi_\mu^2$.
The coupling factor $\phi_\mu$ is essentially determined by the geometry of the constriction, i.e., its width $w_{\rm c}$ and thickness $d$.
$S_\Phi$ depends on the SQUID inductance $L$ and on the junction parameters $I_0$, resistance $R$ and capacitance $C$.
If the constriction could be made not only arbitrarily thin and narrow, but also arbitrarily short, one could envision a scenario, where $\phi_\mu$ reaches a value around $0.5\,\Phi_0/\mu_{\rm B}$ \cite{Bouchiat09}, while, at the same time, the inductance of the constriction remains small ($\Phi_0$ is the magnetic flux quantum).
Then, $S_\Phi$ could be optimized independently by proper choice of the SQUID size and the junction properties.
For the type of device we discuss here, this is certainly not the case and we thus look for an optimization, which is compatible with technological limitations.
A large coupling $\phi_\mu$ demands an as narrow and thin as possible constriction.
On the other hand, for a too narrow constriction, given a fixed value of $d$, its inductance $L_{\rm c}$ and thus also the total inductance $L$ of the SQUID may become too large, possibly degrading the flux noise.
This may be counterbalanced by choosing a different film thickness and changing, e.g., the junction width $w_{\rm J}$.

In the following sections, we derive explicit expressions for the dependence of $\phi_\mu$ (Sec.~\ref{sec:CouplingFactor}) and $S_\Phi$ (Sec.~\ref{sec:FluxNoise}) on various geometric and electric SQUID parameters, which then allows us to optimize $S_\mu$ (Sec.~\ref{sec:Optimization}).

\section{Coupling Factor}
\label{sec:CouplingFactor}

We numerically calculate the coupling factor $\phi_\mu=\Phi/\mu$, i.e.~the flux $\Phi$ coupled into the SQUID loop by a point-like particle with magnetic moment $\mu$, using the software package 3D-MLSI.
This routine takes explicitly into account the geometry in the plane of the SQUID loop (cf.~Fig.~\ref{Fig:layout}), and is based on the numerical simulation of the two-dimensional (2D) sheet current density distribution $j_{\rm 2D}(x,y)$ in the SQUID loop, using London theory with $\lambda_{\rm L}$ and $d$ (and hence the effective penetration depth in the thin film limit) as adjustable parameters \cite{Khapaev03}.

\subsection{Methods}
\label{subsec:Methods}

Three different methods, which are briefly described in the following, have been developed to calculate $\phi_\mu$.

\textbf{Method 1:}
With 3D-MLSI we choose an arbitrary value for the total current $J$ circulating around the SQUID hole and calculate the corresponding sheet current density distribution $j_{\rm 2D}(x,y)$ in the SQUID loop.
The resulting $j_{\rm 2D}(x,y)$ is then used to calculate the three-dimensional (3D) magnetic field distribution $\bm B(\bm r)$ generated by $J$.
The coupling factor is then obtained from the relation
%
\begin{equation}
\phi_\mu(\bm r,\widehat{\bm e}_\mu) = -\widehat{\bm e}_\mu \cdot \bm B(\bm r)/J
\label{eq:phi-mu-method1}
\end{equation}
%
which was derived in Ref.~[\onlinecite{Nagel11}].
Here, $\widehat{\bm e}_\mu$ is the unit vector along the direction of the magnetic moment $\bm\mu=\mu\,\widehat{\bm e}_\mu$ at position $\bm r$.
This means that Eq.~(\ref{eq:phi-mu-method1}) provides $\phi_\mu$ for any given position $\bm r$ and orientation $\widehat{\bm e}_\mu$ of a point-like magnetic particle.

To capture variations of $\bm B$ with film thickness $d$, we simply assume that the circulating current $J$ flows within a number $n$ of 2D sheets in the $x$-$y$-plane, stacked equidistantly along the $z$-axis from the upper surface (at $z=0$) to the lower surface (at $z=-d$) of the SQUID loop.
The resulting field $\bm B(\bm r$) is obtained by averaging the individual fields generated by the sheets.

In our earlier work (see Ref.~[\onlinecite{Nagel13}] and references therein) we used $n=2$, which corresponds to a circulating current flow only in the upper and lower surface sheet of the SQUID loop.
This approach works well if $d$ is small enough.
However, if one is interested in the scaling of $\phi_\mu$ with $d$ one should use a larger value for $n$, which provides a better approximation of a homogeneous current density distribution within the entire film thickness in $z$-direction, in particular for relatively large $d$.
Since for YBCO $\lambda_{\rm L} \approx 0.7\,\rm{\mu m}$ along the c-axis (here, the $z$-direction), we expect such a homogenous current distribution along $\widehat{\bm e}_z$ for a technologically reasonable thickness ($d\,\lapprox\,\rm{0.5\,\mu m}$).

\textbf{Method 2:}

The expression for the coupling factor $\phi_\mu$ from Eq.~(\ref{eq:phi-mu-method1}), as used for Method 1 does not take into account modifications of $j_{\rm 2D}(x,y)$ due to the strongly inhomogeneous dipole field in close vicinity to the magnetic particle.
Such a modification, however, may become important when the distance between the point-like dipole and the SQUID surface is smaller than the film thickness $d$.
Within Method 2, we achieve a better description of the near-field regime by calculating (with 3D-MLSI) the fluxoid $\Phi_{\rm{fluxoid}}(\bm r$) in the SQUID loop, which is induced  by a ``quasi-dipole'' (mimicking a small magnetic particle at position $\bm r$) with a magnetic moment of $1\,\mu_{\rm B}$.
With this we obtain $\phi_\mu(\bm r)=\Phi_{\rm fluxoid}(\bm r)/\mu_{\rm B}$.
Such a quasi-dipole can be constructed by a properly adjusted circulating current in a tiny loop placed at position $\bm r$.
However, in this case, the orientation $\widehat{\bm e}_\mu$ of the magnetic moment of the quasi-dipole is now fixed by the design of this tiny loop, implemented in 3D-MLSI, which allows only to construct 2D structures in the $x$-$y$-plane.

\begin{figure}[b]
\includegraphics[width=8.5cm]{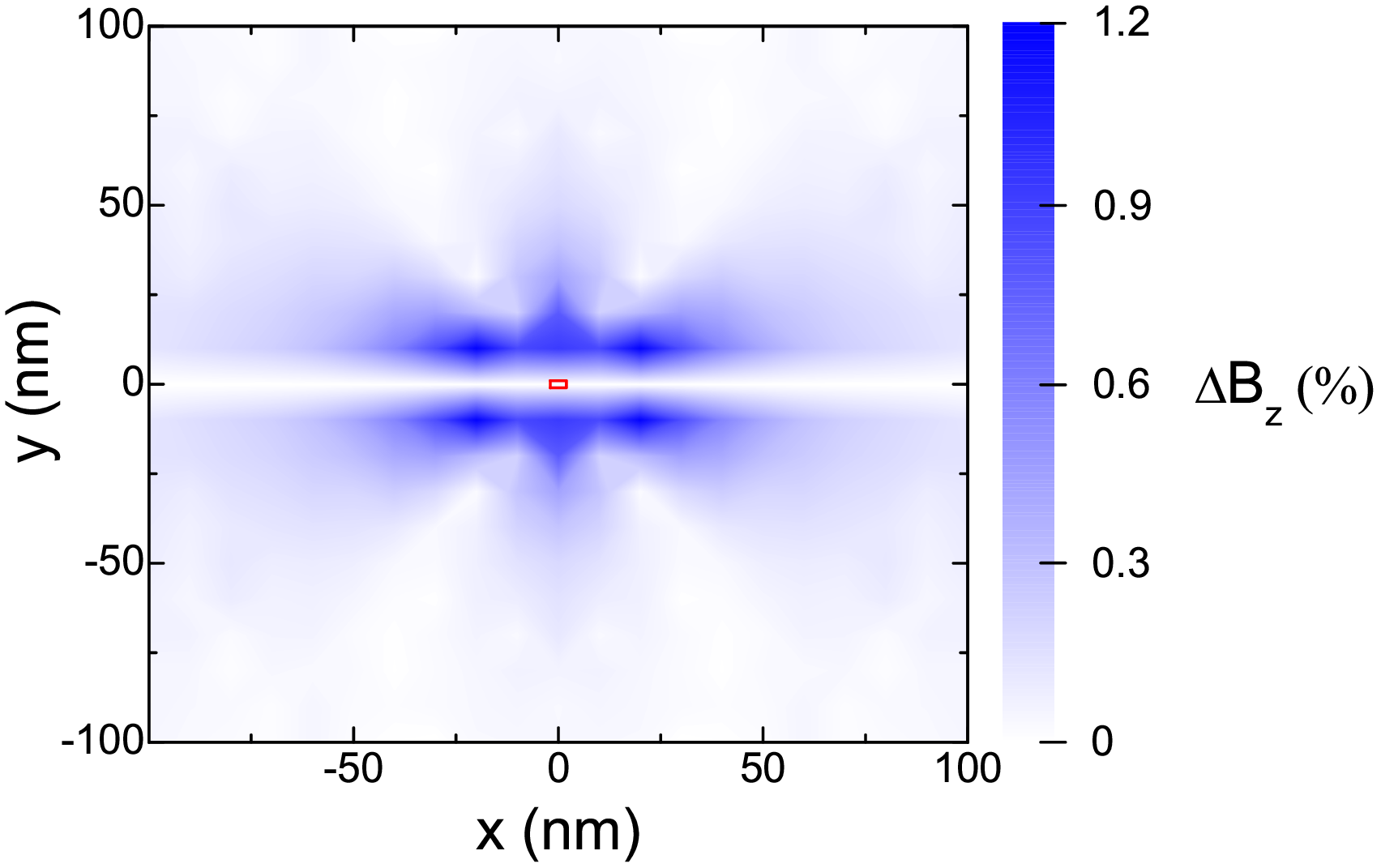}
\caption{Distribution of normalized difference $\Delta B_z(x,y$) in the $z$-components of the quasi-dipole vs ideal dipole fields at $z=0$, with both dipoles centered at $(0,0,z_0=10\,\rm{nm})$.
The red rectangle indicates size and position of the two strips (stacked on top of each other) forming the quasi-dipole.}
\label{Fig:dipoles}
\end{figure}

For instance a quasi-dipole with its magnetic moment oriented along the $z$-axis (i.e.~$\widehat{\bm e}_\mu=\widehat{\bm e}_z$) can be realized by a current circulating in a tiny ring in the $x$-$y$-plane.
Due to the layout of the nanoSQUID considered in this work, it is however more favorable to construct a dipole with magnetic moment pointing in $y$-direction.
Unfortunately, it is not possible to build a corresponding ring within 3D-MLSI.
Instead, we consider two strips (2D current sheets) lying on top of each other with separation $\Delta z = 3$\,nm along the $z$-axis.
Both strips expand 4\,nm and 2\,nm in $x$- and $y$-direction, respectively.
Currents flowing along $\widehat{\bm e}_x$ ($-\widehat{\bm e}_x$) in the upper (lower) strip create a quasi-dipole field with a magnetic moment oriented along $\widehat{\bm e}_y$.
The currents were adjusted to generate the magnetic field distribution of a single $\mu_{\rm B}$.
Furthermore the two strips are regarded as normal conductors by setting $\lambda_{\rm L} \rightarrow \infty$.
The quasi-dipole does not provide the field distribution of an ideal dipole (from a point-like particle) since the two strips are not connected.
However the field generated by the missing links should be of minor relevance since it neither interacts with the superconducting structure nor with the SQUID hole.
In Fig.~\ref{Fig:dipoles} we plot the relative deviation $\Delta B_z$ between the $z$-component of the magnetic field $B_{z,\rm{qd}}$ created by the quasi-dipole and $B_{z,\rm{d}}$ of an ideal dipole
%
\begin{equation}
\Delta B_z=\left|\frac{B_{z,\rm{qd}}-B_{z,\rm{d}}}{B_{z,\rm{d}}}\right|
\label{eq:B_z}
\end{equation}
%
in the $x$-$y$-plane at $z=0$, with both dipoles centered at $\bm r_0=(0,0,z_0=10\,\rm{nm})$ and with an orientation of their magnetic moment along the $y$-axis.
As expected, the quasi-dipole is a very good approximation to an ideal magnetic dipole in the far field regime.
In the near field regime one finds minor deviations of $\Delta B_{z,\rm{max}} \approx 1.2\,\%$, which presumably arise from the finite volume of the quasi-dipole.

For the nanoSQUID structure, the effect of (ideal) flux focussing is taken into consideration by setting the net current $J$ circulating around the hole to zero.
The calculation is deployed for $n=11$ current sheets and the resulting fluxoids are averaged in a similar way as for Method 1.

\textbf{Method 3:}

For this method we again examine the interaction of the quasi-dipole with the SQUID loop.
In contrast to Method 2, (ideal) screening is taken into consideration by setting the fluxoid in the loop to zero.
In other words, a circulating current $J$ is induced in the loop, which counterbalances the coupled flux of the quasi-dipole, due to the diamagnetic response of the SQUID.
The coupling factor is obtained by computing $L$ of the bare SQUID within 3D-MLSI and calculating $\phi_\mu(\bm r)=\Phi_{\rm fluxoid}(\bm r)/\mu_{\rm B}=LJ/\mu_{\rm B}$.
As before, the calculation is performed for $n=11$ current sheets.

\subsection{Comparison of methods}
\label{subsec:comparison}

To compare the three methods, we calculate $\phi_\mu$ for a particle with its magnetic moment oriented along $\hat{e}_y$, which corresponds to the optimum direction of the applied external magnetic field for our SQUID design.
In all cases, we find a maximum in $\phi_\mu(\bm r$) if the dipole is placed as close as possible on top of the constriction at its center in the $x$-$y$-plane.
For the following considerations, we set the origin of our coordinate system at the center of the constriction in the $x$-$y$-plane at the upper surface of the superconducting film.

\begin{figure}[t]
\includegraphics[width=8cm]{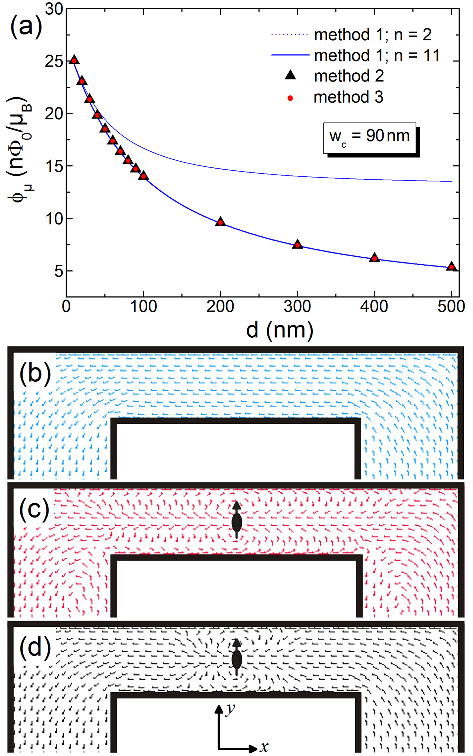}
\caption{Comparison of methods used for calculating the coupling factor and current distribution in a $w_{\rm c}=90\,\rm{nm}$ wide constriction (for $\lambda_{\rm L}=250\,$nm).
(a) $\phi_\mu(d)$ for a particle at $z_0$=10\,nm; position and direction of magnetic moment is indicated in (c) and (d).
(b)--(d) 3D-MLSI output of the current distribution in the $x$-$y$-plane calculated with (b) Method 1 (identical distribution for all $n$ sheets), (c) Method 2 and (d) Method 3 (for uppermost sheet at $z=0$).
Arrows indicate the local direction of currents.}
\label{Fig:methods}
\end{figure}

Assuming that the particle is placed at the position $\bm r_0=(0,0,z_0)$ with $z_0=10\,$nm above the constriction (without an Au layer, which can be removed without affecting the junction properties), we calculate $\phi_\mu(d)$ in the range 10\,nm $\le d \le$ 500\,nm for the three presented methods [cf.~Fig.~\ref{Fig:methods}(a)].

For Method 1, with $n=2$ current sheets, $\phi_\mu (d)$ saturates for $d\,\gapprox\,200\,$nm to $\phi_{\mu,\rm{s}} \approx \frac{1}{2} \phi_\mu (d=10\,\rm{nm})$.
Since the current $J$ is circulating in sheets at the lower ($z=-d$) and upper ($z=0$) surface of the superconductor, the field $B_y(z_0=10\,\rm{nm})$ induced by the lower sheet decays as $d$ increases.
However the field induced by the upper sheet remains constant and thus the mean value of $B_y$ as well, as soon as the contribution from the lower sheet becomes negligible for large enough $d$.
Obviously, the saturation in $\phi_\mu (d)$ is an artefact stemming from the simple approximation of the current distribution along $\widehat{\bm e}_z$ by the currents in only two surface sheets.

Turning to Method 1 with $n=11$ current sheets, the unphysical saturation of $\phi_\mu (d)$ is eliminated.
Similar calculations with $n=101$ and $n=1001$ reveal the same behavior of $\phi_\mu (d)$ for the range of thickness shown.
As expected, Method 1 with $n=2$ and $n=11$ yields the same $\phi_\mu (d)$ for very small $d$.

Albeit Method 1 provides a sensible approximation of $\phi_\mu$ for currents flowing across the entire film thickness if $n$ is large enough, it does not incorporate the effect of local screening currents induced by a magnetic particle in close proximity to the SQUID.
This becomes obvious by comparison of the current distributions in the region of the constriction, as shown for Method 1 in Fig.~\ref{Fig:methods}(b) and for Methods 2 and 3 in Figs.~\ref{Fig:methods}(c) and \ref{Fig:methods}(d), respectively.
The latter two feature a more complex current distribution, arising from local screening currents.
The corresponding dependence $\phi_\mu (d)$ for Method 2 and 3 [cf.~Fig.~\ref{Fig:methods}(a)], however, show qualitatively and quantitatively the same behavior as for Method 1 (with $n=11$).
Accordingly, the local screening currents taken into account in Method 2 and 3 do not alter $\phi_\mu$ in the near field regime as compared to Method 1.

Concluding this section, we have shown that all three methods constitute a valid approach for calculating the coupling factor, since each technique gives the same dependence  $\phi_\mu(d, w_{\rm c})$ for large enough values of $n$.

\subsection{Results}
\label{subsec:results}

\begin{table*}[t]
\begin{tabular}{c c c c c c c c c c c c c c}\hline\hline
$\lambda_{\rm L}$   & $\phi_{\mu,0}$            & $d_0$ & $w_0$ & $w_{\rm c}'$  & $L'$          & $L_{\rm e}'$  & $L_{\rm b}'$  & $L_{\rm b}''$ & $b$   & r     & $L'/d_0$  & $S_{\Phi,0}^{1/2}$        & $S_{\mu,0}^{1/2}$             \\
(nm)                & (n$\Phi_0/\mu_{\rm B}$)   & (nm)  & (nm)  & (nm)          & (pH$\cdot$nm) & (pH$\cdot$nm) & (pH$\cdot$nm) & (pH$\cdot$nm) &       &       & (pH)      &(n$\Phi_0/{\rm Hz}^{1/2}$) & ($\mu_{\rm B}/{\rm Hz}^{1/2}$)\\ \hline\hline
250                 & 49                        & 120   & 102   & 7             & 85            & 56            & 25            & 120           & 0.29  & 2.73  & 0.71      & 12.6                      & 0.26                          \\ \hline
335                 & 78                        & 83    & 53    & 4.8           & 143           & 100           & 45            & 150           & 0.31  & 2.45  & 1.72      & 19.7                      & 0.25                          \\ \hline
\end{tabular}
\caption{Summary of fit parameters from numerical simulations on nanoSQUIDs for two different values of $\lambda_{\rm L}$.
The values for $S_{\Phi,0}^{1/2}$ and $S_{\mu,0}^{1/2}$ are given for $T=4.2\,$K and $I_0R=0.5\,$mV.}
\label{tab:fit-parameters}
\end{table*}

As already mentioned in Sec.~\ref{sec:nanoSQUIDdesign}, the coupling factor should also depend on the width of the constriction.
Hence, we computed $\phi_\mu$ in the range $10\,{\rm nm} \le w_{\rm c} \le 500\,$nm and 10\,nm$\,\le d \le\,$500\,nm, assuming that the quasi-dipole is placed $10\,$nm above the center of the constriction, as in the previous section.
The numerical results can be approximated by
%
\begin{equation}
\phi_\mu (d, w_{\rm c}) \approx \frac{\phi_{\mu,0}}{(1+\frac{d}{d_0}) (1+\frac{w_{\rm c}}{w_0})}\;,
\label{eq:phi_mu}
\end{equation}
%
with the values for the fitting parameters $\phi_{\mu,0}$, $d_0$ and $w_0$ given in Tab.~\ref{tab:fit-parameters} for two different values of $\lambda_{\rm L}$.
As expected, $\phi_\mu$ decreases with increasing width $w_{\rm c}$ and thickness $d$.
Within the simulation range, we find a monotonic decrease of $\phi_\mu(d, w_{\rm c})$, with a slightly weaker decay in $\phi_\mu(d)$ as for $\phi_\mu(w_{\rm c})$.

By modifying the distance $z_0$ between magnetic particle and the upper surface of the superconductor, we find qualitatively the same dependence as in Eq.~(\ref{eq:phi_mu}) within $10\,\rm{nm} \leq z_0 \leq 1000\,\rm{nm}$ with absolute values scaling like $\phi_\mu(z_0) \propto z_0^{-3/2}$.
Since the optimization of $\phi_\mu$ does only trivially depend on the distance between particle and SQUID, we can absorb $\phi_\mu(z_0)$ into $\phi_{\mu, 0}$.

\section{Flux Noise}
\label{sec:FluxNoise}

To determine the flux noise of the SQUID in the thermal white noise regime, we use the theoretical expression obtained from Langevin simulations
%
\begin{equation}
S_\Phi = f(\beta_L) \Phi_ 0 k_{\rm B} T L / {I_0 R},
\label{eq:S_Phi}	
\end{equation}
%
which is valid for a Stewart-McCumber parameter $\beta_{\rm C} \equiv 2\pi I_0 R^2 C/\Phi_0\,\lapprox\,1$ and $\Gamma\beta_L<0.1$ \cite{Chesca-SHB-2}.
Here, $\Gamma \equiv 2\pi k_{\rm B} T / I_0 \Phi_0$ is the noise parameter, and $\beta_L \equiv 2 L I_0/\Phi_0$ is the screening parameter.
For $\beta_L > 0.4$, $f(\beta_L) \approx 4(1+\beta_L)$.
For lower values of $\beta_L$, $S_\Phi$ increases.

The first factor to be discussed is $I_0 R$.
The junction resistance $R$ can be varied to some extent by varying the thickness $d_{\rm Au}$ of the Au layer covering the YBCO film; the maximum achievable value is the unshunted junction normal state resistance $R_{\rm N}$ (for $d_{\rm Au}=0$).
For $24^\circ$ YBCO grain boundary junctions, $I_0 R_{\rm N}$ values $\sim2-3\,\rm{mV}$ are achievable at $4.2\,\rm{K}$ \cite{Hilgenkamp02}.
However, such junctions typically have hysteretic IVCs.
We thus demand $\beta_{\rm C}\,\lapprox\,1$ to avoid hysteresis.
Ideally, one would like to derive an expression for $I_0 R$ as a function of $w_{\rm J}$, $d$ and $d_{\rm Au}$ using the constraint $\beta_{\rm C}\,\lapprox\,1$ and assuming certain values for the critical current density $j_0$, unshunted normal junction resistance times area $\rho \equiv R_{\rm N} w_{\rm J} d$ and capacitance per junction area $C'$.
However, the scaling of $R$ with $w_{\rm J}$, $d$ and $d_{\rm Au}$ is currently not known.
Furthermore, an estimate of $C'$ as a function of $w_{\rm J}$ and $d$, based on various scaling laws available in literature \cite{Moeckly95,McBrien00,Navacerrada05} is quite difficult, in particular since it is difficult to determine $C$ for underdamped YBCO GBJs and since the stray capacitance due to the commonly used $\rm{SrTiO_3}$ substrates may play an important role \cite{Beck96}.
On the other hand, we have fabricated nanoSQUIDs from $24^\circ$ YBCO GBJs with different junction widths $85\le w_{\rm J} \le 440\,$nm and film thicknesses 50, 100 and 300\,nm, using the focused ion beam (FIB) milling technique as described in Ref.~[\onlinecite{Schwarz13}].
Parameters of some of those devices are listed in Tab.~\ref{tab:device-parameters}.
Except for the devices with both, small film thickness ($d=50\,$nm) and narrow junctions ($w_{\rm J}\approx 100\,$nm), which tend to have slightly lower $I_0R$ and $j_0$, typical values for our devices are $I_0R\approx 0.5\,$mV and $j_0=3-5\,{\rm mA}/\mu{\rm m}^2$ at $T=4.2\,$K.
Below we will find an optimum junction width well above 100\,nm and a very weak dependence of the optimum spin sensitivity on film thickness for $100\,{\rm nm}\,\lapprox\, d \,\lapprox\, 500\,$nm.
Thus, rather than introducing an ill-defined scaling of $I_0R$ with $w_{\rm J}$ and $d$, below we fix $I_0 R=0.5\,\rm{mV}$ and $j_0=3\,{\rm mA}/\mu{\rm m}^2$ as realistic values.

We next determine the dependence of the SQUID inductance $L$ on the various geometrical parameters.
We separate the SQUID into the constriction (inductance $L_{\rm c}$, length $l_{\rm c}$, width $w_{\rm c}$), the two (symmetric) bridges containing the junctions (inductance $L_{\rm J}$, length $l_{\rm J}$, width $w_{\rm J}$), the two corners connecting the constriction and the junction arms (inductance $L_{\rm e}$), and the bottom part of the SQUID (inductance $L_{\rm b}$), as indicated in Fig.~\ref{Fig:layout}.
Then, $L$ is given by
%
\begin{equation}
L=L_{\rm c}+2L_{\rm J}+2L_{\rm e}+L_{\rm b}\;.
\label{eq:L}
\end{equation}
%
We should find $L_{\rm c} (w_{\rm c}, l_{\rm c}, d)$, $L_{\rm J} (w_{\rm J}, l_{\rm J}, d)$, $L_{\rm e} (w_{\rm c}, w_{\rm J}, d)$ and $L_{\rm b} (l_{\rm c}, w_{\rm J}, d)$.
From 3D-MLSI simulations we find the parametrization $L_{\rm c} (w_{\rm c}, l_{\rm c}, d) \approx L' \cdot l_{\rm c}/{w_{\rm c} d}$.
This expression fits the computed $L_{\rm c}$ well, within the parameter range $10\,{\rm nm} \le l_{\rm c},w_{\rm c},d \le 500\,$nm, covered by the simulations.
We use the same parametrization for $L_{\rm J} (w_{\rm J}, l_{\rm J}, d)$.
For the corners we find, within a 15\,\% variation with respect to $w_{\rm J}$ and $w_{\rm c}$, the expression $L_{\rm e} \approx L_{\rm e}'/d$.
Finally, we find $L_{\rm b} \approx L_{\rm b}' l_{\rm c}/w_{\rm J} d + L_{\rm b}''/d$.
The fitting parameters $L'$, $L_{\rm e}'$, $L_{\rm b}'$ and $L_{\rm b}''$ are summarized in Tab.~\ref{tab:fit-parameters} for two different values of $\lambda_{\rm L}$.
Inserting these expressions into Eq.~(\ref{eq:L}) yields
%
\begin{equation}
L \approx \frac{L'}{d} \left\{ \frac{l_{\rm c}}{w_{\rm c}} + \frac{2l_{\rm J}+b l_{\rm c}}{w_{\rm J}}+r \right\}\;,
\label{eq:L2}
\end{equation}
%
with $b \equiv L_{\rm b}'/L'$ and $r \equiv (2L_{\rm e}'+L_{\rm b}'')/L'$ (cf.~Tab.~\ref{tab:fit-parameters}).
We note that in our simulations we have adjusted $\lambda_{\rm L}=250\,$nm to be consistent with most of the experimentally determined values of $L$ for our nanoSQUIDs.
This value is consistent with the literature on $\lambda_{\rm L}$ in the $a$-$b$-plane of epitaxially grown $c$-axis oriented YBCO thin films\cite{Zaitsev02,Arpaia14}.
However, for some devices we find good agreement between measured and simulated values of $L$ only if we assume larger values for $\lambda_{\rm L}$, e.g.~$\lambda_{\rm L}=335\,$nm for ''exp.~device 1a`` listed in Tab.~\ref{tab:device-parameters}.

For the minimization of $S_\mu$, we will use $\beta_L$ as a variable parameter.
Since both, $L$ and $w_{\rm J}$ are not independent of each other and are related to $\beta_L$, we express both as functions of $\beta_L$.
This will allow us to eliminate $L$ and $w_{\rm J}$ in the final expression for $S_\mu$ which has to be optimized.
With $\beta_L = 2 I_0 L/\Phi_0$ and $I_0 = j_0 w_{\rm J} d$, we obtain
\begin{equation}
w_{\rm J} (\beta_L, L)=\frac{\Phi_0 \beta_L}{2 j_0 d L}\;.
\label{eq:w_J}
\end{equation}
%
Inserting this into Eq.~(\ref{eq:L2}) yields
%
\begin{equation}
L(\beta_L) \approx \frac{L'}{d} \left( \frac{l_{\rm c}}{w_{\rm c}}+r \right) \left\{1-\frac{\kappa}{\beta_L}\right\}^{-1}\;,
\label{eq:L3}
\end{equation}
%
with
\begin{equation}
\kappa(l_{\rm J}, l_{\rm c}, j_0) \equiv 2(2 l_{\rm J}+b l_{\rm c})j_0 L'/\Phi_0\;.
\label{eq:kappa}
\end{equation}
%
Inserting Eq.~(\ref{eq:L3}) into Eq.~(\ref{eq:S_Phi}) and using $f(\beta_L) = 4(1+\beta_L)$ finally yields
%
\begin{equation}
S_\Phi(d,w_{\rm c},\beta_L) \approx S_{\Phi,0} \frac{d_0}{d} \left(\frac{l_{\rm c}}{w_{\rm c}}+r \right) \frac{1+\beta_L}{1-\frac{\kappa}{\beta_L}}\;,
\label{eq:S_Phi2}
\end{equation}
%
with $S_{\Phi,0}^{1/2} \equiv 2 \sqrt{\frac{\Phi_0 k_{\rm B} T L'}{I_0 R d_0}}$ (cf.~Tab.~\ref{tab:fit-parameters}).
The most important result here is the scaling $S_\Phi \propto 1/d$.
This is due to the fact that the SQUID inductance $L \propto 1/d$ within the simulation range for $d$, because of the increase of the kinetic inductance contribution with decreasing $d$ below $\lambda_{\rm L}$.
For $d \; \gapprox \; 2 \lambda_{\rm L}$ we expect a saturation of $L(d)$ and hence of $S_\Phi(d)$.
However, we will neglect this for the optimization of $S_\mu$, since values for $d\,\gapprox\,500\,$nm are outside the simulation range and since we cannot expect to produce high-quality GBJs for such large values of $d$.

\section{Optimization of Spin Sensitivity via improved SQUID geometry}
\label{sec:Optimization}

With Eqs.~(\ref{eq:phi_mu}) and (\ref{eq:S_Phi2}) we find the spin sensitivity $S_\mu^{1/2}=S_\Phi^{1/2}/\phi_\mu$.
The individual dependencies on $d$, $\beta_L$ and constriction parameters $w_{\rm c}$ and $l_{\rm c}$ can be separated.
Hence, we can express the spin sensitivity as
%
\begin{equation}
S_\mu^{1/2}(d,w_{\rm c},\beta_L) = S_{\mu,0}^{1/2} \cdot s_d(d)\cdot s_{\beta_L}(\beta_L) \cdot s_{\rm c}(w_{\rm c},l_{\rm c}) \;,
\label{eq:S_mu}
\end{equation}
%
with $S_{\mu,0}^{1/2} \equiv S_{\Phi,0}^{1/2}/\phi_{\mu,0}$ (cf.~Tab.~\ref{tab:fit-parameters}) and with
%
\begin{equation}
s_d(d) \equiv \sqrt{\frac{d_0}{d}} + \sqrt{\frac{d}{d_0}}\;,
\label{eq:s_d}
\end{equation}
\begin{equation}
s_{\beta_L}(\beta_L) \equiv \sqrt{\frac{1+\beta_L}{1-\frac{\kappa}{\beta_L}}}\;,
\label{eq:s_betaL}
\end{equation}
\begin{equation}
s_{\rm c}(w_{\rm c},l_{\rm c}) \equiv \left(1 + \frac{w_{\rm c}}{w_0}\right) \sqrt{\frac{l_{\rm c}}{w_{\rm c}}+r}\;.
\label{eq:s_wc}
\end{equation}
%
\begin{figure}[b]
\includegraphics[width=8.5cm]{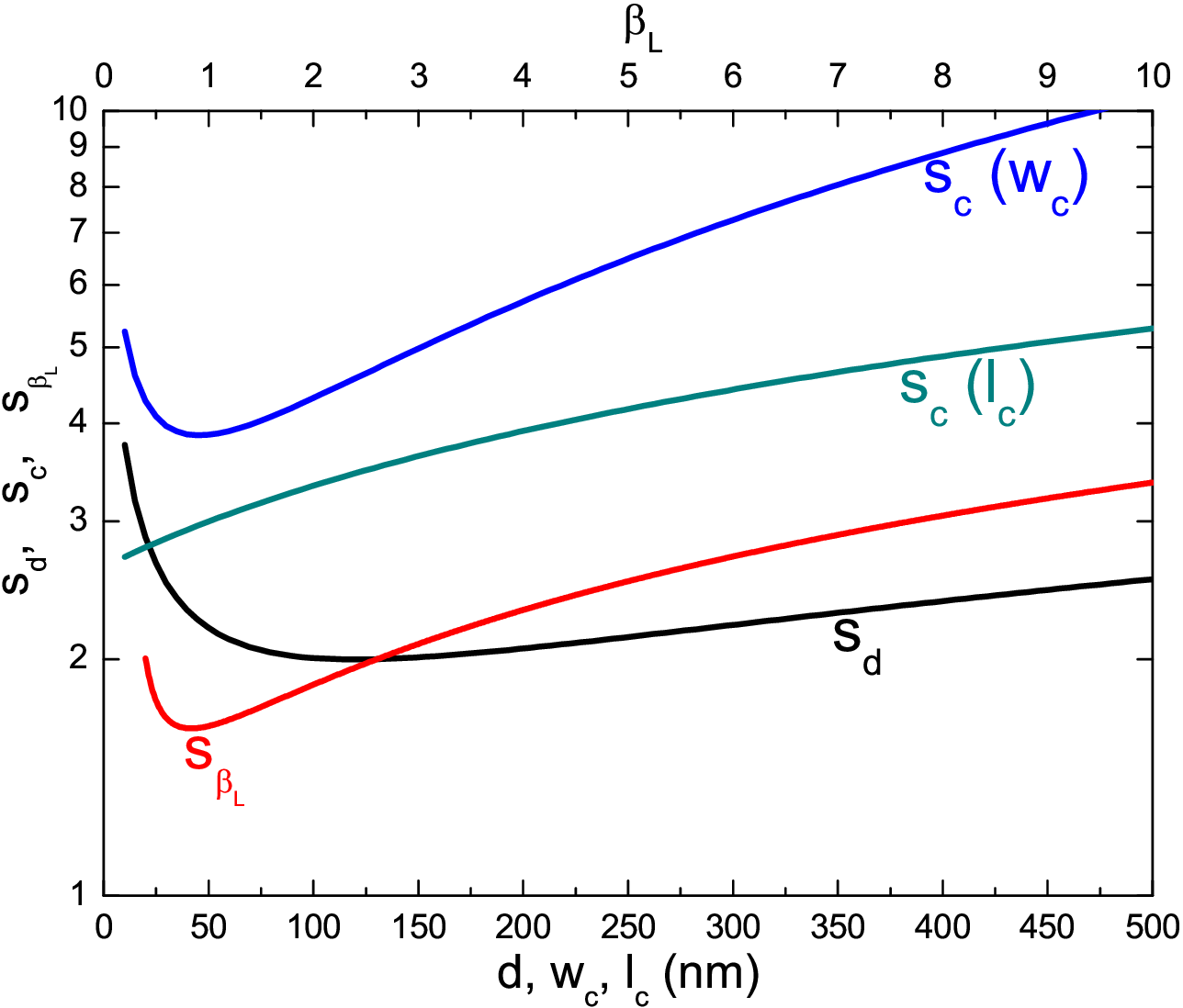}
\caption{Scaling of the terms $s_d(d)$, $s_{\beta_L}(\beta_L)$ for $\kappa=0.26$,  $s_{\rm c}(w_{\rm c})$ for $l_{\rm c}=200\,$nm and $s_{\rm c}(l_{\rm c})$ for $w_{\rm c}=60\,$nm, which enter the spin sensitivity in Eq.~(\ref{eq:S_mu}) as calculated from Eqs.~(\ref{eq:s_d})--(\ref{eq:s_wc}) with $\lambda_{\rm L}=250\,$nm.}
\label{Fig:SQUID_opt}
\end{figure}
%
Figure \ref{Fig:SQUID_opt} shows $s_d(d)$, $s_{\beta_L}(\beta_L)$ for fixed $\kappa$, and $s_{\rm c}(w_{\rm c})$ and $s_{\rm c}(l_{\rm c})$ for fixed $l_{\rm c}$ and $w_{\rm c}$, respectively, for $\lambda_{\rm L}=250\,$nm.
In the following we discuss the optimum choice of the various parameters.

For $s_d(d)$ from Eq.~(\ref{eq:s_d}) we obtain a shallow minimum at $d_{\rm min}=d_0$, and a rather weak dependence for $d\,\gapprox\,100\,$nm.
This indicates that with increasing $d$ above $\sim100$\,nm the decrease in kinetic inductance (and hence in flux noise) and coupling factor almost compensate each other within the simulation range.
Hence, the optimization of the spin sensitivity with respect to film thickness is straightforward, although, the proper choice of $d$ is not very crucial as long as $d\gapprox 100\,$nm. However, in order to avoid too large aspect ratios $d/w_{\rm c}$ and $d/w_{\rm J}$, it is advisable to fix the optimum film thickness to $d_{\rm opt}=d_{\rm min}$.
This in turn fixes the optimum value for $s_d$ according to Eq.~(\ref{eq:s_d}) to
\begin{equation}
s_{d,\rm opt}=s_d(d_{\rm min})=2\;.
\label{eq:d_opt}
\end{equation}

The evaluation of Eq.~(\ref{eq:s_betaL}) shows a much more pronounced dependence for $s_{\beta_L}(\beta_L)$ with a clear minimum at $\beta_{L,\rm{min}}=\kappa(1+\sqrt{1+\kappa^{-1}})$, and $s_{\beta_L}(\beta_{L,\rm{min}})=\sqrt{\kappa}+\sqrt{\kappa +1}$.
For $\kappa=0.26$ used in Fig.~\ref{Fig:SQUID_opt}, we obtain $\beta_{L,\rm{min}}\approx 0.83$ and $s_{\beta_L}(\beta_{L,\rm{min}})\approx 1.6$.
Both, $\beta_{L,\rm{min}}(\kappa)$ and $s_{\beta_L}(\beta_{L,\rm{min}})$ decrease monotonically with decreasing $\kappa$, which implies that $\kappa$ should be as small as possible.
However, as mentioned above, for $\beta_L<0.4$ the flux noise increases again with further decreasing $\beta_L$, and Eq.~(\ref{eq:s_betaL}) is not applicable.
Hence, the optimum value for $\beta_L$ is $\beta_{L,\rm{opt}}=0.4$, which then fixes the optimum value for $\kappa$ via the relation $\beta_{L,\rm{min}}(\kappa)$ to
\begin{equation}
\kappa_{\rm opt}=\frac{\beta_{L,\rm{opt}}^2}{1+2\beta_{L,\rm{opt}}}=\frac{4}{45}\approx 0.09\;.
\label{eq:kappa_opt}
\end{equation}
%
Accordingly, the optimum value for $s_{\beta_L}$ in Eq.~(\ref{eq:s_betaL}) yields
\begin{equation}
s_{\beta_L,\rm{opt}}=s_{\beta_L}(\beta_{L,\rm{opt}},\kappa_{\rm opt})=\frac{3}{\sqrt{5}}\approx 1.3\;.
\label{eq:betaL_opt}
\end{equation}
%
We note that according to Eq.~(\ref{eq:kappa}), the choice of $\kappa=\kappa_{\rm opt}$ relates the optimum length $l_{\rm{J,opt}}$ of the bridges containing the GBJs and $l_{\rm c}$ via
%
\begin{equation}
l_{\rm{J,opt}}=\frac{\kappa_{\rm opt}\Phi_0}{4j_0L'}-\frac{b}{2}l_{\rm c}\;.
\label{eq:l_Jvsl_c}
\end{equation}
%
Since $b/2\approx 0.15\ll 1$, the dependence $l_{\rm{J,opt}}(l_{\rm c})$ is quite weak.
For our choice of $j_0=3\,{\rm mA}/\mu{\rm m}^2$ and with $\lambda_{\rm L}=250\,$nm, Eq.~(\ref{eq:l_Jvsl_c}) yields $l_{\rm{J,opt}}\approx 180\,{\rm nm}-0.15 l_{\rm c}$, i.e. $l_{\rm{J,opt}}$ decreases only slightly from $\sim 180\,$nm to $\sim 150\,$nm for $l_{\rm c}=0$ to 200\,nm.
Hence, the choice of $l_{\rm c}$ (together with $j_0$ and $\lambda_{\rm L}$) fixes $l_{\rm{J,opt}}$.

By inserting $d=d_{\rm{opt}}=d_0$, $\beta_L=\beta_{L,\rm{opt}}$ and $\kappa=\kappa_{\rm opt}$ into Eq.~(\ref{eq:L3}), we find for the optimized SQUID inductance
\begin{equation}
L_{\rm opt}\approx 1.3\,\frac{L'}{d_0}\left(r+\frac{l_{\rm c}}{w_{\rm c}}\right)\;,
\label{eq:L_opt}
\end{equation}
%
i.e.~$L_{\rm opt}\approx 2.5\,{\rm pH}+0.91\,{\rm pH}\cdot\frac{l_{\rm c}}{w_{\rm c}}$ for $\lambda_{\rm L}=250\,$nm and roughly a factor of two larger values for $\lambda_{\rm L}=335\,$nm.
Inserting this into Eq.~(\ref{eq:w_J}), we find for the optimum junction width
\begin{equation}
w_{\rm{J,opt}}=\frac{7\Phi_0}{45L'j_0}\;\frac{1}{r+\frac{l_{\rm c}}{w_{\rm c}}}\;.
\label{eq:w_Jopt}
\end{equation}
%
For our choice of $j_0=3\,{\rm mA}/\mu{\rm m}^2$, the prefactor in Eq.~(\ref{eq:w_Jopt}) is $\approx 1.26\,\mu{\rm m}$ (750\,nm) for $\lambda_{\rm L}=250\;(335)\,$nm; i.e.~the optimum junction width decreases monotonically with increasing ratio $l_{\rm c}/w_{\rm c}$ from $\sim 340\;(270)\,$nm for $l_{\rm c}/w_{\rm c}=1$ to $\sim 100\;(60)\,$nm for $l_{\rm c}/w_{\rm c}=10$, with $\lambda_{\rm L}=250\;(335)\,$nm.

\begin{figure}[b]
\includegraphics[width=8.5cm]{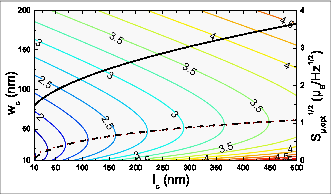}
\caption{Contour plot of optimized spin sensitivity $S_{\mu,\rm{opt}}^{1/2}(l_{\rm c},w_{\rm c})$ (for $T=4.2\,$K, $I_0R=0.5\,$mV, $d=120\,$nm and $\beta_L=0.4$).
Numbers at contour lines are in units of $\mu_{\rm B}/{\rm Hz}^{1/2}$.
Dashed and dotted lines show $w_{\rm{c,min}}(l_{\rm c})$ from Eq.~(\ref{eq:w_cmin}) and approximation by power law dependence, respectively.
The solid black line shows $S_{\mu,\rm{opt}}^{1/2}(l_{\rm c})$ for $w_{\rm c}=w_{\rm{c,min}}$.
All quantities were calculated for $\lambda_{\rm L}=250\,$nm.}
\label{Fig:S-mu-contour}
\end{figure}

Finally, as shown in Fig.~\ref{Fig:SQUID_opt}, the relation $s_{\rm c}(w_{\rm c},l_{\rm c})$, given by Eq.~(\ref{eq:s_wc}) yields a monotonic decrease of $s_{\rm c}$ with decreasing $l_{\rm c}$ and a clear minimum in $s_{\rm c}(w_{\rm c})$ at
%
\begin{equation}
w_{\rm{c,min}}=\frac{l_{\rm c}}{4r}\left(\sqrt{1+\frac{8rw_0}{l_{\rm c}}}-1\right)\;,
\label{eq:w_cmin}
\end{equation}
%
which can be approximated by a power law dependence $w_{\rm{c,min}}\approx w_{\rm c}'\cdot(l_{\rm c}/{\rm nm})^{0.35}$ (cf.~dashed and dotted lines in Fig.~\ref{Fig:S-mu-contour}) with $w_{\rm c}'=7\;(4.8)\,$nm for $\lambda_{\rm L}=250\;(335)\,$nm.
Accordingly, $s_{\rm c}$ can be minimized by choosing $w_{\rm c}=w_{\rm{c,min}}(l_{\rm c})$.
This yields
\begin{equation}
s_{\rm{c,opt}}(l_{\rm c})=\left\{1+\frac{w_{\rm c}'}{w_0}\left(\frac{l_{\rm c}}{\rm nm}\right)^{0.35}\right\}\;\sqrt{r+\frac{\rm nm}{w_{\rm c}'}\left(\frac{l_{\rm c}}{\rm nm}\right)^{0.65}}\;.
\label{eq:s_copt}
\end{equation}
%
Both, $w_{\rm{c,min}}(l_{\rm c})$ and $s_{\rm{c,opt}}(l_{\rm c})$ decrease monotonically with decreasing $l_{\rm c}$.
This implies that $l_{\rm c}$ should be made as small as possible.

\begin{table*}[t]
\begin{tabular}{c c c c c c c c c c c c c c c c c c c}\hline\hline
           &$d$    & $l_{\rm c}$ & $l_{\rm J}$  & $w_{\rm c}$   & $w_{\rm J}$   & $\beta_L$ & $L$       & $I_0$     & $R$       & $I_0 R$   & $j_0$         & $L_{\rm c}$   & $L_{\rm J}$   & $L_{\rm e}$   & $L_{\rm b}$   & $S_\Phi^{1/2}$            & $\phi_\mu$            & $S_\mu^{1/2}$\\
units    & nm      & nm          & nm           & nm            & nm            &           & pH        & $\mu$A    & $\Omega$  & mV        & mA/$\mu$m$^2$ & pH            & pH            & pH            & pH            & n$\Phi_0/\rm{Hz}^{1/2}$   & n$\Phi_0/\mu_{\rm B}$ & $\mu_{\rm B}/\rm{Hz}^{1/2}$\\ \hline\hline
opt.     &&&&&&&&&&&&&&&&&&\\
device1  &\rb{120} &\rb{44}      &\rb{174}      &\rb{25}        &\rb{280}       &\rb{0.40}  &\rb{4.1}   &\rb{101}   &\rb{5.0}   &\rb{0.5}   &\rb{3}         &\rb{1.3}       &\rb{0.44}      &\rb{0.47}      &\rb{1.0}       &\rb{36}                    &\rb{20}                &\rb{1.8}\\ \hline
opt.     &&&&&&&&&&&&&&&&&&\\
device2  &\rb{120} &\rb{100}     &\rb{200}      &\rb{60}        &\rb{316}       &\rb{0.45}  &\rb{4.1}   &\rb{114}   &\rb{4.4}   &\rb{0.5}   &\rb{3}         &\rb{1.2}       &\rb{0.45}      &\rb{0.47}      &\rb{1.1}       &\rb{36}                    &\rb{15}                &\rb{2.4}\\ \hline
exp.     &\lb{50}  &\lb{300}     &\lb{400}      &\lb{90}        &\lb{130}       &\lb{0.65}  & 36        &\lb{18.5}  &\lb{7.0}   &\lb{0.13}  &\lb{2.85}      &\lb{5.7}       &\lb{5.2}       &\lb{1.1}       &\lb{3.6}       & 1300                      &\lb{18}                & 71\\
device1a &         &             &              &               &               &           & (22)      &           &           &           &               &               &               &               &               & (228)                     &                       & (12)\\ \hline
exp.     &\lb{50}  &\lb{535}     &\lb{435}      &\lb{50}        &\lb{85}        &\lb{1.29}  & 42        &\lb{31.4}  &\lb{10.2}  &\lb{0.32}  &\lb{7.39}      &\lb{18}        &\lb{8.7}       &\lb{1.1}       &\lb{5.6}       & 600                       &\lb{23}                & 26\\
device1b &         &             &              &               &               &           & (43)      &           &           &           &               &               &               &               &               & (185)                     &                       & (8.0)\\ \hline
exp.     &\lb{100} &\lb{500}     &\lb{500}      &\lb{420}       &\lb{190}       &\lb{0.78}  & 8.9       &\lb{91}    &\lb{5.4}   &\lb{0.49}  &\lb{4.79}      &\lb{1.0}       &\lb{2.2}       &\lb{0.56}      &\lb{1.9}       & 450                       &\lb{5.2}               & 86\\
device2a &         &             &              &               &               &           & (8.5)     &           &           &           &               &               &               &               &               & (60)                      &                       & (11)\\ \hline
exp.     &\lb{100} &\lb{475}     &\lb{455}      &\lb{410}       &\lb{140}       &\lb{1.37}  & 9.1       &\lb{155}   &\lb{3.1}   &\lb{0.47}  &\lb{11}        &\lb{0.98}      &\lb{2.8}       &\lb{0.56}      &\lb{2.0}       & 400                       &\lb{5.3}               & 75\\
device2b &         &             &              &               &               &           & (9.7)     &           &           &           &               &               &               &               &               & (72)                      &                       & (13)\\ \hline
exp.     &\lb{300} &\lb{300}     &\lb{450}      &\lb{120}       &\lb{280}       &\lb{0.87}  & 2.9       &\lb{315}   &\lb{1.4}   &\lb{0.44}  &\lb{3.75}      &\lb{0.71}      &\lb{0.46}      &\lb{0.19}      &\lb{0.49}      & 240                       &\lb{6.4}               & 37\\
device3a &         &             &              &               &               &           & (2.5)     &           &           &           &               &               &               &               &               & (37)                      &                       & (5.7)\\ \hline
exp.     &\lb{300} &\lb{485}     &\lb{480}      &\lb{195}       &\lb{285}       &\lb{1.01}  & 2.2       &\lb{471}   &\lb{1.7}   &\lb{0.78}  &\lb{5.51}      &\lb{0.70}      &\lb{0.48}      &\lb{0.19}      &\lb{0.54}      & $<240$                    &\lb{4.8}               & $<50$\\
device3b &         &             &              &               &               &           & (2.6)     &           &           &           &               &               &               &               &               & (25)                      &                       & (5.3)\\ \hline
\end{tabular}
\caption{Summary of geometric and electric nanoSQUID parameters (as defined in the text).
The values for ``opt.~device 1'' are calculated for optimized parameters obtained for a given constriction length $l_{\rm c}$, with $\lambda_{\rm L}=250\,$nm.
For ``opt.~device 2'' we used more relaxed values for $w_{\rm c}$, $l_{\rm c}$ and $l_{\rm J}$ and otherwise identical input parameters for $d$, $j_0$, $I_0 R$, $\lambda_{\rm L}$ with correspondingly optimized $\beta_L$ and adjusted $w_{\rm J}$.
For the experimental devices we quote experimentally determined values for $L$ and $S_\Phi^{1/2}$  together with values (in brackets) which are calculated with Eqs.~(\ref{eq:L2}) and (\ref{eq:S_Phi}), respectively, with $\lambda_{\rm L}=250\,$nm.
Here, the flux noise was calculated based on the measured SQUID inductance $L$.
Accordingly, the values in brackets for the spin sensitivity $S_\mu^{1/2}$ are based on the calculated values for the flux noise $S_\Phi^{1/2}$.}
\label{tab:device-parameters}
\end{table*}

All numbers in the following paragraph are quoted for $\lambda_{\rm L}=250\,$nm.
For $l_{\rm c}=500\,$nm we find $w_{\rm{c,min}}\approx 60\,$, which is feasible to realize with our FIB technology; however upon shrinking $l_{\rm c}$ it becomes increasingly hard to realize devices with optimum constriction width $w_{\rm{c,min}}(l_{\rm c})$.
Fortunately, it turns out that the degradation in spin sensitivity is not very severe if $w_{\rm c}$ deviates from $w_{\rm{c,min}}$, as long as one can keep $w_{\rm c}$ below, say, 100\,nm.
This is illustrated in the contour plot in Fig.~\ref{Fig:S-mu-contour}, which shows the spin sensitivity for optimized $d$ and $\beta_L$, i.e.~$S_{\mu,\rm{opt}}^{1/2}(l_{\rm c},w_{\rm c})=S_{\mu,0}^{1/2}\cdot s_{d,\rm opt}\cdot s_{\beta_L,{\rm opt}}\cdot s_{\rm c}(l_{\rm c},w_{\rm c})\approx 0.69\,\mu_{\rm B}/{\rm Hz}^{1/2}\cdot s_{\rm c}(l_{\rm c},w_{\rm c})$ for $T=4.2\,$ and $I_0R=0.5\,$mV.
Within the plotted range, the spin sensitivity lies in most cases between 2 and 4\,$\mu_{\rm B}/{\rm Hz}^{1/2}$, and practically for an optimized device the spin sensitivity is limited by both, the smallest length and linewidth which can be realized for the constriction.
The solid line in Fig.~\ref{Fig:S-mu-contour} shows $s_{\rm{c,opt}}(l_{\rm c})$ according to Eq.~(\ref{eq:s_copt}), i.e.~with the additional condition $w_{\rm c}=w_{\rm{c,min}}(l_{\rm c})$.
If we take $l_{\rm c}=44\,$nm, corresponding to $w_{\rm{c,min}}=25\,$nm as the current limitation for our FIB patterning technology, we calculate $S_{\Phi,\rm{opt}}^{1/2} \approx 36\,\rm{n \Phi_0/Hz^{1/2}}$ and $\phi_{\mu,\rm{opt}} \approx 20\,\rm{n \Phi_0/\mu_{\rm B}}$, giving an optimized spin sensitivity $S_{\mu,\rm{opt}}^{1/2} \approx 1.8\,\rm{\mu_{\rm B}/Hz^{1/2}}$.
Corresponding SQUID parameters are listed in Tab.~\ref{tab:device-parameters} (``opt.~device 1'').
If we take more easily achievable values $w_{\rm c}=60\,$nm, $l_{\rm c}=100\,$nm and $l_{\rm J}=200\,$nm  (other input parameters are the same as for the initial optimization), we still get $S_\mu^{1/2}=2.4\,\rm{\mu_{\rm B}/Hz^{1/2}}$ (see Tab.~\ref{tab:device-parameters} for parameters of ``opt.~device 2'').

\section{Discussion}
\label{sec:Discussion}

In the following, we discuss some practical issues regarding the realization of optimized YBCO GBJ nanoSQUIDs.
The optimization of the spin sensitivity given by Eq.~(\ref{eq:S_mu}) certainly depends on the control over the various input parameters, which are not always known precisely.
For example, $I_0R$ and $j_0$ of YBCO GBJs can vary significantly, even on the same chip\cite{Hilgenkamp02}, and sometimes we find values for $\lambda_{\rm L}$ significantly above 250\,nm.

Starting with the prefactor $S_{\mu,0}^{1/2}$, this depends on $T$ and $I_0R$.
Regarding operation temperature $T$, this will certainly depend on the different applications the nanoSQUIDs will be used for.
Hence, this is not a parameter which should be used for optimization.
Still, the use of YBCO SQUIDs based on GBJs offers operation from close to their transition temperature $T_{\rm c}$ (say, 77\,K) down to the mK regime.
The very large range of operation temperatures is certainly a significant advantage over nanoSQUIDs based on other materials or other junction types such as constriction junctions, which often can only be operated in a very limited temperature interval.
The $I_0R$ product does only enter into the expression for the spin sensitivity via $S_{\mu,0}\propto 1/I_0R$.
Hence, any variation in $I_0R$ does not affect the optimization of the device geometry.
Obviously, as large as possible values for $I_0R$ are helpful for improving the spin sensitivity.

The term for $s_d$ depends on the film thickness $d$ only, and due to the shallow minimum in $s_d(d)$, slight deviations from $d=d_{\rm opt}=120\,$nm (for $\lambda_{\rm L}=250\,$nm) or larger values for $\lambda_{\rm L}$ will have an almost negligible effect on $S_\mu^{1/2}$.

The term for $s_{\rm c}$ depends only on the geometry of the constriction and on $\lambda_{\rm L}$.
Here, technological limitations imposed by the patterning technique and possible edge damage effects are crucial, since the smallest achievable $s_{\rm c}$ will depend on the smallest achievable length $l_{\rm c}$ and width $w_{\rm c}$ of the constriction.
For our FIB patterning technique, we currently do not know what the final limits for the minimum achievable values for $l_{\rm c}$ and $w_{\rm c}$ are, and how strong edge damage effects are.
Further investigations are required to determine (and reduce) edge damage effects, which will finally limit the minimum achievable constriction size.

The term $s_{\beta_L}$ depends on $\beta_L$ and $\kappa$.
Here, $j_0$ enters into the optimization only via $\kappa\propto j_0$.
A variation in $j_0$ will modify the optimum length $l_{\rm{J,opt}}(j_0,l_{\rm c})$ [cf.~Eq.~(\ref{eq:l_Jvsl_c})] and width $w_{\rm{J,opt}}\propto 1/j_0$ [cf.~Eq.~(\ref{eq:w_Jopt})], which are required for maintaining $\beta_L\approx 0.4$ (and hence $s_{\beta_L}=s_{\beta_L,\rm{opt}}$).
Fortunately, $j_0$ can be measured prior to FIB patterning, which allows to adjust the geometry of the bridges straddling the GBJs.
Hence, as long as $j_0$ does not change significantly after FIB milling\cite{Nagel11}, and as long as the conditions for $l_{\rm{J,opt}}$ and $w_{\rm{J,opt}}$ can be fulfilled, the optimized spin sensitivity is not affected by variations in $j_0$.

A variation in $\lambda_{\rm L}$ has a similar effect as a variation in $j_0$, since $\kappa\propto L'$ and $L'$ increases with $\lambda_{\rm L}$ (cf.~Tab.~\ref{tab:fit-parameters}).
However, it is difficult to determine $\lambda_{\rm L}$ prior to FIB patterning in order to adjust $w_{\rm J}$ and $l_{\rm J}$ properly.
For fixed geometrical parameters, we find that an increase in $\lambda_{\rm L}$ from 250 to 335\,nm decreases the coupling factor only very slightly, as long as $w_{\rm c}\lapprox 100\,$nm.
The strongest effect comes from the increase in $L'$ by a factor of $\sim 1.7$, which increases $L$ and $\beta_L$, which both enter into the flux noise.
Depending on the value of $\beta_L$, this induces an increase in $S_\Phi^{1/2}$ (and in $S_\mu^{1/2}$) by a factor of approximately 1.4 to 1.7.

Finally, we would like to comment on two additional practical issues.
First, the predicted optimized spin sensitivity around a few $\mu_{\rm B}/{\rm Hz}^{1/2}$ is in particular due to the reduction in SQUID inductance for an optimized geometry, yielding improved flux noise.
However, we should mention that for YBCO SQUIDs the measured flux noise is often significantly higher than the theoretically predicted one\cite{Koelle99}.
For the experimental devices listed in Tab.~\ref{tab:device-parameters} the measured $S_\Phi^{1/2}$ was a factor 3.2 to 7.5 higher than predicted by Eq.~(\ref{eq:S_Phi}).
Hence, we expect the predicted spin sensitivities to be too low by a similar factor if compared with experimental results.

Second, the optimization procedure as described in this work is based on calculating the white thermal noise of the SQUIDs.
However, it is well known that $I_0$ fluctuations can lead to a flux noise $S_\Phi$ which scales with the measurement frequency $f$ as $1/f^\alpha$ with $\alpha$ typically close to 1, and it is also known that for YBCO GBJs such a $1/f$ noise contribution can be quite large\cite{Koelle99}.
For YBCO nanoSQUIDs with improved white thermal noise around $100\,{\rm n}\Phi_0/{\rm Hz}^{1/2}$ and below, this implies that the $1/f$ noise may dominate at frequencies up to the MHz range.
Hence, in order to utilize the full potential of such SQUIDs, the implementation of bias reversal schemes for suppression of $1/f$ noise from $I_0$ fluctuations will be very important.
Furthermore, for dc SQUIDs based on metallic superconductors such as Nb, it has been shown that below $T\approx 1\,$K additional sources of low-frequency excess flux noise may become important, which cannot be eliminated by bias reversal\cite{Wellstood87} (for more recent work see e.g.~[\onlinecite{Choi09,Drung11}] and references therein).
In YBCO nanoSQUIDs also similar effects may be present and deserve further studies.

\section{Conclusions}
\label{sec:Conclusions}

In summary, we have performed a detailed analysis of the coupling factor $\phi_\mu$ and the spectral density of flux noise $S_\Phi$, and hence of the spin sensitivity $S_\mu^{1/2}=S_\Phi^{1/2}/\phi_\mu$ for grain boundary junction dc nanoSQUIDs.
Based on the calculation of $\phi_\mu$ and $S_\Phi$, we derived an explicit expression for the spin sensitivity $S_\mu^{1/2}$ as a function of the geometric and electrical parameters of our devices.
This allows for an optimization of $S_\mu^{1/2}$, which predicts a spin sensitivity of a few $\mu_{\rm B}/\rm{Hz}^{1/2}$.
Such a low value for $S_\mu^{1/2}$ can be achieved by realization of very low inductance nanoSQUIDs with ultra-low flux noise on the order of $100\,\rm{n \Phi_0/Hz^{1/2}}$ or even below, in the thermal white noise regime.
This poses severe challenges on proper readout electronics for such SQUIDs.
It remains to be shown whether or not the readout of such ultralow-noise SQUIDs is feasible and whether or not the envisaged values for the spin sensitivity can also be achieved in high fields, which is a major driving force for using these grain boundary junction nanoSQUIDs.

\acknowledgments

J.~Nagel and T.~Schwarz acknowledge support by the Carl-Zeiss-Stiftung.
We gratefully acknowledge fruitful discussions with D.~Drung.
This work was funded by the Nachwuchswissenschaftlerprogramm of the Universit\"{a}t T\"{u}bingen, and by the Deutsche Forschungsgemeinschaft (DFG) via projects KO 1303/13-1 and SFB/TRR 21 C2.



\end{document}